# Getting Science Beyond the Research Community: Examples of Education and Outreach from the IceCube Project

J. Madsen
*UWRF, River Falls, WI 54022, USA*
*For the IceCube Collaboration*[1]

The IceCube collaboration has built an in-ice neutrino telescope and a surface detector array, IceTop, at the South Pole. Over 5000 digital optical modules have been deployed in a cubic kilometer of ice between 1450 and 2450 m below the surface. The novel observatory provides a new window to explore the universe. The combination of cutting-edge discovery science and the exotic Antarctic environment is an ideal vehicle to excite and engage a wide audience. Examples of how the international IceCube Collaboration has brought the Universe to a broader audience via the South Pole are described.

## 1. Introduction

The international IceCube collaboration has recently completed a multipurpose neutrino and cosmic ray observatory located on the Amundsen Scott Base at the South Pole[2]. After six seasons of construction, the biggest science project ever attempted in Antarctica and one of the largest detectors in the world is providing a new window to view the Universe. The allure of cutting-edge discovery science combined with the exotic Antarctic environment and international partners provide multiple opportunities to excite and engage a wide audience. This proceeding describes examples of the IceCube collaboration's education and outreach efforts targeted for students, teachers and the general public.

## 2. IceCube

The motivation for the IceCube project was to realize the dream of building a cubic-kilometer scale neutrino telescope to explore the Universe with neutrino messengers. The Antarctic Muon and Neutrino Detector Array (AMANDA) established the feasibility of using the nearly 3000 m thick ice at the South Pole as medium for detecting neutrinos. A hot water drill was used to melt holes 60 cm in diameter and up to 2450 m deep. Photomultipliers embedded in the ice detected the light resulting from charged particles produced from neutrino interactions in and near the instrumented volume. The novelty of the idea was recognized in 1999 by Scientific American when AMANDA was named the weirdest of the seven wonders of Modern Astronomy [1].

Construction on the IceCube detector began during the 2003-2004 austral summer season. The short construction season annually brings a pulse of activity that offers unique opportunities for real and virtual participation. The Amundsen Scott Station opens around the end of October each year, and the last return flight of the season leaves in mid-February. During construction, IceCube personnel were a significant fraction of the approximately 150 to 200 South Pole population. Typically there were about thirty drillers, with a significant fraction of the drill team returning for most of the six seasons. In addition, there were about 20 other IceCube personnel at the South Pole including scientists, engineers, IT personnel, graduate students and postdoctoral researches, and occasionally, undergraduate students and high school teachers. Two or three IceCube winterovers remain at the South Pole Station to maintain the IceCube observatory during the cold, dark, winter months.

### 2.1. Collaboration

The IceCube Collaboration consists of over 250 scientists from 39 institutions, about half of whom are in the United States with the balance from Australia, Barbados, Belgium, Canada, Germany, Great Britain, Japan, New Zealand, Sweden, and Switzerland. A map of collaborating institutions and a list of the funding agencies are shown in Figure 1. The large international representation, together with the remote location of the observatory, provides opportunities for real and virtual participation on a variety of levels at locations around the world. Examples of these experiences will be provided below.

---

[1] http:/icecube.wisc.edu/collaboration/authors
[2] http://www.icecube.wisc.edu/news/current

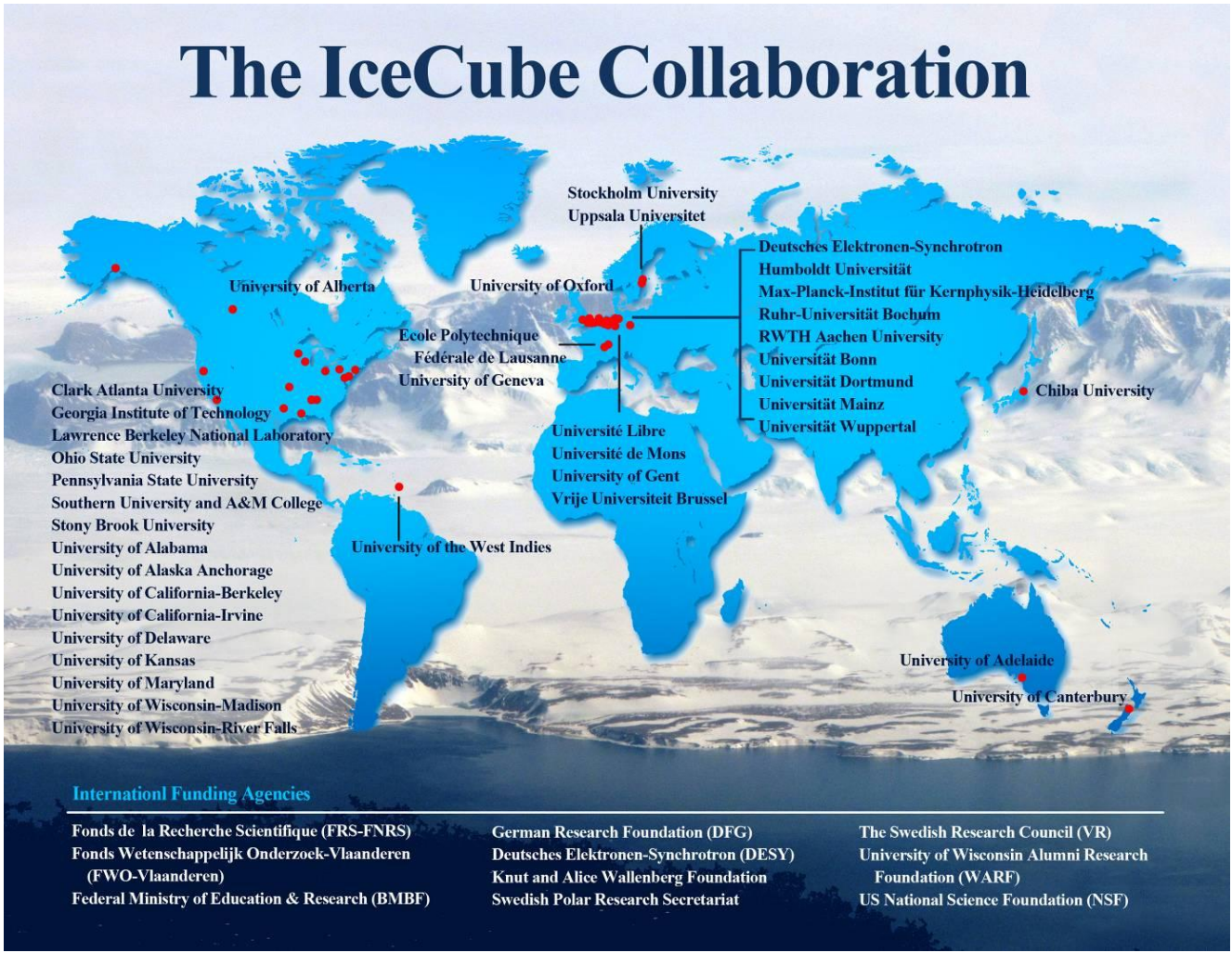

Figure 1: IceCube Collaboration institutions and construction funding agencies[3].

### 2.2. Observatory

The IceCube observatory consists of an in-ice array of Digital Optical Modules (DOMs) that are deployed in holes drilled with hot water. A DOM consists of a 10" photomultiplier tube housed inside a glass pressure vessel with a data acquisition system and light emitting diodes (leds) for calibration purposes (see Figure 2b). A cable with sixty DOMs is lowered into the water filled hole that freezes, locking the DOMs in place. The DOMs plus the cable in a given hole is referred to as a string. There are 78 strings on a triangular grid with a 125 m separation with DOMs deployed between 1450 and 2450 m below the surface. A sub array of 8 strings at the center of the detector, known as DeepCore has more closely spaced strings and DOMs deployed in the deepest clearest ice. DeepCore lowers the energy threshold for detecting neutrinos and, together with the outer strings, makes it possible to extend the solid angle of detection for low energy neutrinos to $4\pi$.

There is also a surface array known as IceTop that contains 81 stations on the same 125 m triangular grid as the original in-ice array. Each consists of pairs of ice Cherenkov tanks approximately 1.6 m in diameter and 0.9 m deep that are monitored by two DOMs, one configured for high gain and the other for low gain to extend the dynamic range of the tanks. The distance between tanks in a station is 10 meters. A diagram of the IceCube Observatory is shown in Figure 2a.

---

[3] Image by Jamie Yang and Mark Krasberg

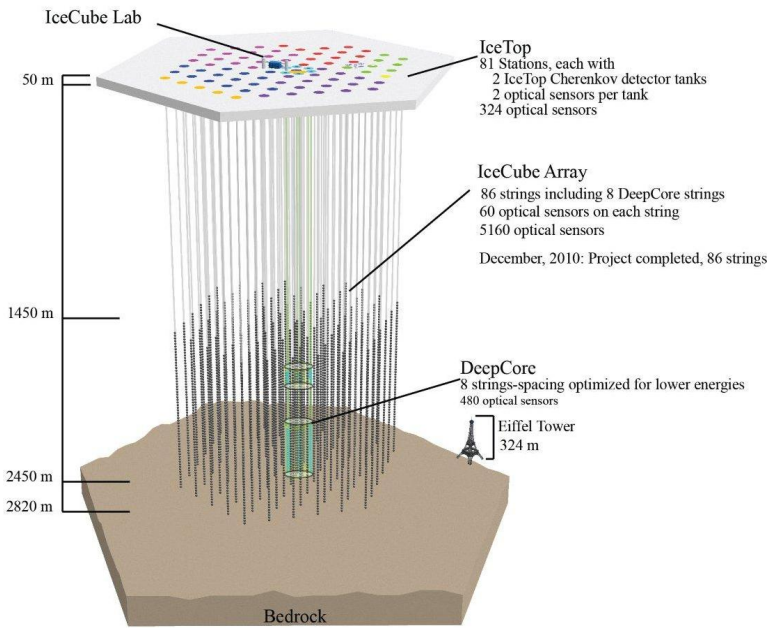
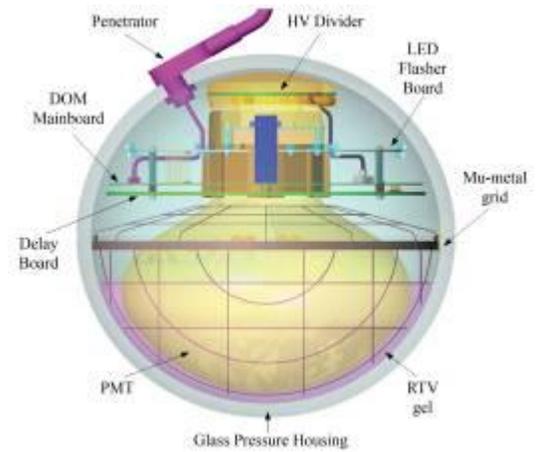

Figure 2a (left) provides a diagram of the IceCube observatory[4]. Figure 2b (right) shows the main parts of a DOM[5].

### 2.3. Science

The primary focus of the IceCube project, as originally proposed, was to study the universe using high energy neutrinos. The discovery potential of new facilities is one of the captivating aspects of science for the general public. Another facet is the possibility to shed light on long-standing mysteries like the:

- origin and acceleration mechanism of the highest energy cosmic rays [2]
- source of gamma ray bursts [3]
- composition of dark matter [4]

IceCube is poised to make contributions to solving these puzzles and others as the full capabilities of the observatory are developed. For example, a dust logger devised to characterize the instrumented ice has enabled IceCube researchers to study glacial dust layers and explore their connections to past climatological events [5]. By comparing dust logger results from multiple drill holes, it may be possible to determine wind speeds tens of thousands of years ago [6]. IceTop, originally designed for calibration and veto purposes or the neutrino telescope and cosmic ray studies above $10^{15}$ eV, now also operates in a mode capable of detecting low energy particles produced during solar storms [7]. Examples of how IceCube brings this science to the broader community are described in the next section.

### 3. Education and Outreach

The goal of the IceCube education and outreach efforts is to get science beyond the research community. The fascination with the extreme Antarctic location and the novelty of new observational approach provide hooks to excite and engage the public, students, and teachers. The result is opportunities to demonstrate that science is an on-going pursuit to understand the Universe rather than the organized collection of established results, as may typically be presented. Significant education and outreach efforts are on-going throughout the IceCube collaboration. Specific examples of efforts in a few general categories based on the depth, duration, and intended audience will be discussed next.

---

[4] Image by Jamie Yang
[5] Image by IceCube Collaboration

### 3.1. One-time events

IceCube collaboration members have given hundreds of presentations to public and school groups. In addition to the compelling science, the fascination with living and working conditions in the extreme Antarctic environment are a big draw. A number of successful displays have also been developed to illustrate the operation of the observatory. This is possible because the observatory is relatively simple, especially when compared to traditional particle physics detectors. There is really only one active component, the DOM. The in-ice neutrino telescope consists of a grid of DOMs imbedded in the ice. To convey the essence of the IceCube neutrino telescope, DOMs can be suspended in space to illustrate the structure (Figure 3a).

In addition to traditional general interest talks, IceCube collaborators used targeted approaches to reach broader audiences. A micro scale hot water drilling activity (Figure 3b) has proven popular, especially with elementary school age children. The inaugural Nuclear Science Day[6] at Lawrence Berkeley National Laboratories drew over 150 boy and girl scouts. Among other activities, the scouts heard a talk about cosmic rays from an IceCube scientist and received career information. IceCube collaborators have also shown a willingness to engage older learners by presenting informal public talks at local taverns. The University of Wisconsin-River Falls (UWRF) visiting professor program has supported an IceCube scientist on two occasions. During their two days on the UWRF campus, they provided guest lectures on their work, delivered an evening public talk, and were generally available to meet with interested students one-on-one. One visit included a highly popular exhibit of the visiting professor's art work, providing another way to attract an audience that would not typically attend science events.

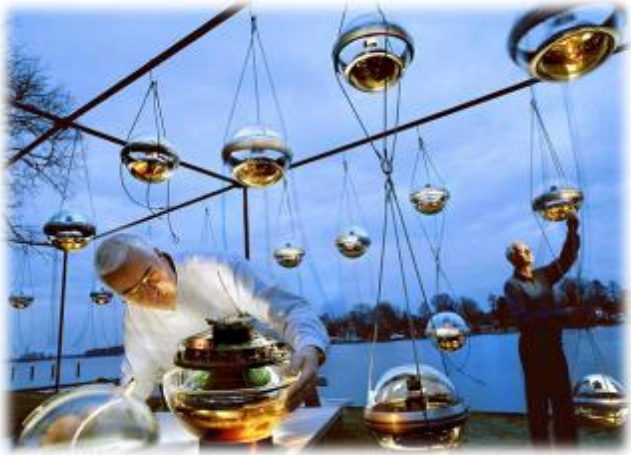 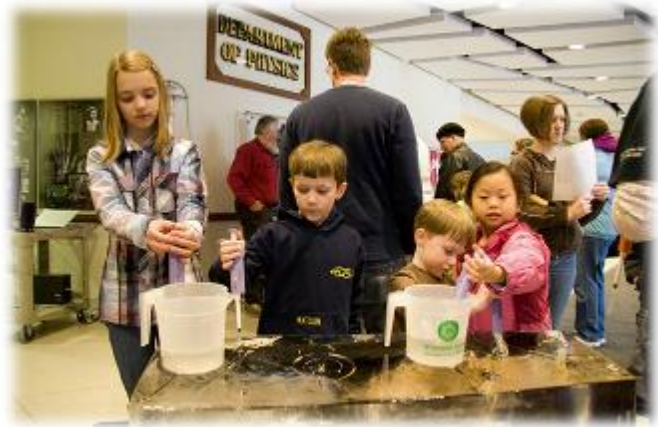

Figure 3a (left) shows preparations for a public event at DESY-Zeuthen in Germany[7].
Figure 3b (right) shows the hot water drilling activity[8].

### 3.2. Field Experiences for students and teachers

The extreme Antarctic environment at the South Pole is one of the main draws for interest in the IceCube project. The IceCube collaboration continues to build on the tradition established by the AMANDA collaboration to offer opportunities for undergraduate students and teachers to participate in IceCube research. Many IceCube institutions have ongoing undergraduate student projects that have made significant contributions. In this section, a few examples of field experiences for undergraduates and teachers will be described.

---

[6] http:/nuclearscienceday11.lbl.gov/
[7] Photo by Peter Ginter
[8] Photo by Laurel Bacque

The AMANDA and IceCube projects have had five teachers work at the South Pole in the last decade. One teacher, Matts Petterson, was from Sweden; the others, Jason Petula, Eric Muhs, Casey O'Hara, and Katey Shirey, were from the USA. Because of population constraints, these teachers needed to fully contribute while at the South Pole. In addition, the goal was to communicate their experiences as broadly as possible. Training for the on-ice research responsibilities were handled by the AMANDA and IceCube collaborations. Partnerships with NSF supported programs that pair polar researchers with teachers---Teachers Experiencing Antarctica and the Arctic (TEA)9 for the first three teachers, and PolarTREC[10] for the last two----have significantly enhanced the impact of their experiences. Casey and Katey's experience also profited from their association and support from the Knowles Science Teaching Foundation[11].

Students have also befitted from international opportunities within the collaboration. Five UWRF undergraduate students and one student from the two-year college UW-Marathon County have spent the summer doing research at Stockholm University in Sweden over the course of three summers. UWRF, the University of Delaware, and the University of Uppsala also worked on a project to calibrate the response of an IceTop tank to low energy particles. An IceTop tank was constructed in a freezer container in Sweden and placed on the icebreaker Oden as it cruised from Sweden to Antarctica and back. The known latitude dependence of the geomagnetic cutoff, the minimum primary cosmic ray energy needed to reach sea level, was used perform an absolute energy calibration. Three undergraduates traveled on the Oden. Drew Anderson, UWRF physics major, traveled from Sweden to Uruguay. In Uruguay, Samantha Jakel, who started at two-year college UW-Rock County, boarded the icebreaker to complete the trip to Antarctica. She is now studying electrical engineering at UW-Madison. Kyle Jero, a UWRF Physics major, took the last leg from Antarctica to Chile. Their results and experience were presented at a variety of venues[12] including the Committee on Undergraduate's Poster Session at the Capitol Hill symposium in Washington, DC and the 24th National Council of Undergraduate Research Meeting in Montana.

### 3.3. Classroom Enrichment

The University of Maryland, Pennsylvania State University, and UWRF have all run multiple courses for high school teachers and/or students on IceCube science. UWRF has had IceCube programs for high school teachers and/or students for the last dozen summers. For example, the Oden cruise was the theme for the eight day residential summer 2010 science and math program for the UWRF Upward Bound (UB) educational program. Approximately two dozen low-income 9-12 grade students from underrepresented groups were engaged in an innovative, inquiry-based learning experience.

The curriculum was developed and taught by teachers with polar research experience----Eric Muhs, Katey Shirey, and Steve Stevenoski. The UB students started by learning how to represent data with contour maps as a lead into understanding the geomagnetic cutoff data. After spending time creating and mapping their own miniature landforms, they moved on to using GPS units to map out the UWRF outdoor amphitheater. To further connect to the work on the Oden, the UB students' week culminated with a cruise in kayaks down the Kinnickinnic River in River Falls. They collected a multidimensional data set including water temperature, depth, flow rate, and atmospheric pressure at multiple locations that were time and position stamped with GPS units.

### 3.4. Social Media and Site Broadcasts

To expand and diversify the IceCube Education and Outreach efforts, the IceCube project has a presence on Facebook, Twitter and in multiple blogs. According to a Nielsen Company study from August 2010, U.S. internet users spent over twenty percent of their time on social networking sites [8]. These tools are increasingly becoming the default tools for information exchange and it is important to utilize applications with which the target audience is comfortable.

Social media also has additional capabilities that make it more powerful than traditional static web pages offering one-way communication. Social media allows for discussion and feedback in real time and provides a convenient venue to archive postings. During construction of the neutrino detector updates and photos were posted of the progress that brought this phase to life for viewers. The social media tools allow connections to be made between interested people and IceCube personnel around the world. Even the winterovers at the South Pole engage in the conversation from time to time, sending photos and updates from the South Pole. The Facebook page list articles of interest, and is a platform to post photos, advertise events, and communicate with our audience.

---

[9] http://tea.armadaproject.org/, http://tea.armadaproject.org/tea_petulafrontpage.html, http://tea.armadaproject.org/tea_muhsfrontpage.html
[10] http://www.polartrec.com/, http://www.polartrec.com/member/casey-ohara, http://www.polartrec.com/member/katey-shirey
[11] KSTF supports young professionals with technical degrees who want to teach, http://www.kstf.org/
[12] www.cur.org/postersession.html, www.umt.edu/ncur2010/

In each construction season, there have been multiple site broadcasts over the internet from the South Pole. These interactive sessions give the public to hear first-hand from IceCube personnel---scientist, drillers, students, teachers, and support personnel---while they are on the "ice". Due to limited bandwidth and satellite access that restricts the hours available each day, coordinating broadcasts for convenient times around the world is sometimes challenge. However, the broadcasts have proven popular and remain an essential part of the IceCube activities each Antarctic season.

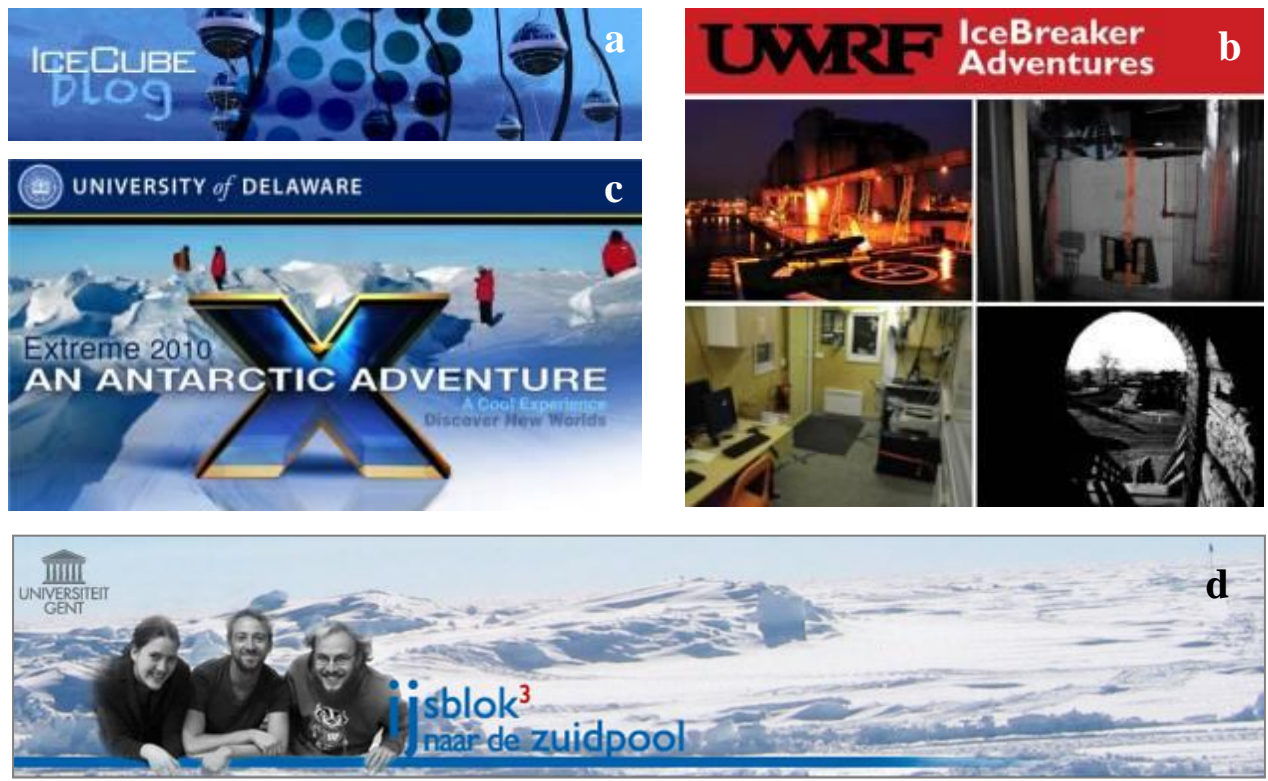

Figure 4: Examples of IceCube blogs[13] at (a) UW-Madison, (b) UW-River Falls, (c) University of Delaware, and (d) University of Gent, Belgium.

## 4. Summary

The construction and operation of the IceCube observatory at the South Pole provides unique opportunities for education and outreach. The IceCube collaboration seeks to engage groups beyond the traditional research community and to allow participation on a variety levels, both virtually and in person. These efforts help explain the process of science, which ultimately helps ensure continued public support while also motivating the next generation of scientists.

### Acknowledgments

We acknowledge support from the following agencies: U.S. NSF—Office of Polar Programs, U.S. NSF—Physics Division, University of Wisconsin Alumni Research Foundation, the GLOW and OSG grids; U.S. DOE, NERSCC, the LONI grid; NSERC, Canada; Swedish Research Council, Swedish Polar Research Secretariat,SNIC, K. and A. Wallenberg

---

[13] (a)www.blog.icecube.wisc.edu, (b) www2.uwrf.edu/icecube/icebreaker.htm, (c) www.expeditions.udel.edu/antarctica/blog-dec-1-2010.html, (d) www.naardezuidpool.ugent.be/blog.1.html

Foundation, Sweden; German Ministry for Education and Research, Deutsche Forschungsgemeinschaft; FSR, FWO Odysseus, IWT, BELSPO, Belgium; Marsden Fund, New Zealand; JSPS, Japan; SNSF, Switzerland. A. Groß is supported by the EU Marie Curie OIF Program, J. P. R. by the Capes Foundation, Brazil, and N.W. by the NSF GRFP. The Oden cruise was supported by NSF—OPP grants 0838838 and 0838534. Support for undergrdaute students in Sweden was provided by a supplemental award to NSF—OPP grant 0636875.